\documentclass[twocolumn,showpacs,aps,amsmath,amssymb,nofootinbib]{revtex4-1}

\usepackage{color}   %option added by cyw eg \pagecolor{blue} \color{red}
\usepackage{graphicx}% Include figure files
\usepackage{dcolumn}% Align table columns on decimal point
\usepackage{bm}% bold math
\usepackage{flushend}

\begin{document}

\title{Multivacuum States in a Fermionic Gap Equation with massive gluons
and confinement}

\author{R. M. Capdevilla}

\affiliation{Physics Department, University of Notre Dame, South Bend, IN, United
States}

\affiliation{Instituto de Fisica Teorica, Universidade Estadual Paulista - UNESP,
Sao Paulo, SP, Brasil}
\begin{abstract}
We study the nontrivial solutions of the QCD fermionic gap equation
including the contribution of dynamically massive gluons and the confining
propagator proposed by Cornwall. Without the confining propagator,
in the case of non-running gluon mass ($m_{g}$), we found the multivacuum
solutions (replicas) reported in the literature and we were able to
define limits on $m_{g}$ for dynamical chiral symmetry breaking.
On the other side, when considering the running in the gluon mass
the vacuum replicas are absent in the limits on $m_{g}$ where the
chiral symmetry is broken. In the pure confining sector, the multivacuum
states are always absent so it is said that only one stable solution
for the gap equation is found as claimed in previous analysis using
different approaches. Finally in the case of the complete gap equation
i.e. with both contributions, the vacuum replicas are also absent
in both cases; with constant and with running gluon mass.
\end{abstract}


\maketitle

\section{Introduction}

In Quantum Chromodynamics (QCD) the fundamental degrees of freedom
of the theory are not detected as free objects and the quark self-energy
can drive the appearance of a dynamical mass. These two phenomena
are known as confinement of quarks and gluons and dynamical chiral
symmetry breaking (CSB), respectively. When studied separately, both
phenomena are partially understood: For the latter, the idea is well
accepted that the chiral condensate obtain a nontrivial vacuum expected
value leading to the generation of a non-zero dynamical quark mass.
In this scheme, the (pseudo)Goldstone bosons associated with the breaking
of the continuous symmetry are the pions. One theoretical tool used
to study this process is the fermionic gap equation (FGE) which can
be obtained from the Schwinger-Dyson equations (SDE) for the fermionic
fields \cite{Roberts1994}. In the case of confinement, an order parameter
used to describe the transition from the confined to the deconfined
phase is the vacuum expectation value of the Polyakov loop $L$ \cite{Greensite2003,Polyakov 1978}.
Although this description is well suited only for pure gluons QCD,
there are some modern approaches with the aim of including quarks
\cite{Kondo2010,Sannino2004}.

Despite the advances in the understanding of both phenomena, one of
the actual challenges for a complete description of the nonperturbative
QCD regime is the connection between those important phases of the
IR behavior of QCD. For example, it has been found that the deconfinement
transition and the chiral symmetry restoration occur approximately
at the same temperature for quarks in the fundamental representation
\cite{Bazavov2009 (Tc eq Td),Cheng2006 (Tc eq Td)}, which is different
for the adjoint representation \cite{Engels2005 (Tc dif Td),Belgici2009 (Tc dif Td)}.
The analysis of this behavior has been recently explored \cite{Capdevilla2014}
in the framework of the gap equation with the inclusion of Cornwall's
confining propagator which has been shown to provide a good description
of the discrepancy between the chiral transition of fundamental and
adjoint quarks. Another issue which concerns the relation of confinement
and chiral symmetry breaking is the idea that removal of central vortices,
may or may not impact in the recovery of the chiral symmetry. It was
found that at least for $SU(2)$ this condition is satisfied \cite{Forcrand su2 1999},
however calculations for $SU(3)$ are not yet conclusive \cite{Bowman su3 2011}.

The authors in reference \cite{Bicudo2002}, using a Hamiltonian approach
to QCD in Coulomb gauge, report that the two dimensional QCD possesses
only one possible vacuum state, given by the solution of the mass-gap
equation, while the four-dimensional theory possesses an excited vacuum
replica. Those results and a theoretical framework are explored in
successive works \cite{Nefediev2003,Bicudo2003}. These authors also
suggest that for the pure linearly rising potential, the interaction
is not strong enough to hold any replicas so that ``only one chirally
nonsymmetric solution to the mass-gap equation may exist'' \cite{Bicudo2002}.
Furthermore, the authors of reference \cite{Kun-lun Wang (Roberts)2012}
studied the fermionic gap equation for pure QCD and they argue that
the excited vacuum states are a consequence of the nature of the gap
equation, since it is an integral equation. However, they also show
that this vacuum states do not affect what we know about the hadronic
spectrum. A quite similar analysis is performed in reference \cite{Raya (Roberts)2013}
but for $\mathrm{QED}_{3}$ in which oscillatory solutions are found
for the gap equation, solutions which are characterized by the number
of zeros.

Nowadays the idea that nonperturbative effects can drive massive propagators
for the gauge bosons as suggested by Cornwall \cite{Cornwall1982}
is well accepted, especially because this result has been confirmed
by lattice simulations \cite{Cucchieri(mg latt)2009,Bogolubsky(mg latt)2009}
and modern approaches using the Dyson-Schwinger equations \cite{Binosi(mg sde)2009,Aguilar(mg sde)2008}.
The consequence of the inclusion of massive gluons in the analysis
of chiral symmetry breaking has been well explored, so that it is
known that for the accepted value \cite{Cornwall1982} of the dynamical
gluon mass $m_{g}\approx2\Lambda_{QCD}$ (being $\Lambda_{QCD}$ the
QCD scale), the fermionic gap equation is too weak to allow the CSB
for quarks in the fundamental representation \cite{Natale1997,Haeri1991,Cornwall(gMass)2008}.
Also, the positivity issues discussed in reference \cite{Cornwall2009}
show that $m_{g}>1.2\Lambda_{QCD}$, values for which CSB is not yet
achieved with the standard fermionic gap equation. As a solution to
this issue, Cornwall proposed \cite{Cornwall2011} a modification
of the Mandelstam confining propagator \cite{Maldestam1979} which
behaves like $1/k^{4}$ for $1/(k^{2}+m^{2})^{2}$. Here, $m$ is
a parameter necessary for entropic reasons, i.e. space-time fluctuations
in the Wilson loop, which turns out to be approximately equal to the
dynamical quark mass at zero momentum ($m\approx M$) in order to
allow the formation of massless bound states \cite{Cornwall2011,Cornwall2012}.
With this form, the confining propagator leads us to a satisfactory
phenomenological interquark potential, and it is also free of IR singularities.
This propagator has to be introduced by hand, being a result of vortices
which appear when the gluon acquires a dynamical mass \cite{Cornwall2012,Cornwall1998}.

In this paper we explore more about the inclusion of the confining
propagator proposed by Cornwall in the fermionic gap equation. We
calculate numerical solutions (for different values of the entropic
parameter $m$) to study the connection between confinement and CSB.
We also check whether or not there are multivacuum states for this
equation. In section \ref{sec:Fermionic-gap-equation} we present
the complete gap equation and the parameters in which it depends.
In section \ref{sec:Fermionic-gap-equation mg} we solve the gap equation
for the one-gluon sector and we found limits on $m_{g}$ below which
the CSB occurs. Section \ref{sec:Fermionic-gap-equation mc} presents
a similar analysis for the confining gap equation and section \ref{sec:Fermionic-gap-equation comp}
does the same for the complete gap equation. We finally present our
summary and conclusions.

\section{Fermionic gap equation\label{sec:Fermionic-gap-equation}}

The dynamical quark mass $M\left(p^{2}\right)$ can be obtained from
the so called complete gap equation, which is the integral equation

\begin{equation}
M\left(p^{2}\right)=\int\frac{d^{4}k}{(2\pi)^{4}}\left\{ G_{c}+G_{g}\right\} \frac{M\left(k^{2}\right)}{k^{2}+M^{2}\left(k^{2}\right)},\label{eq:Comp GE}
\end{equation}

with Kernels $G_{c}$ and $G_{g}$. These correspond to the confining
and one-gluon contribution, given by

\begin{equation}
G_{c}(p-k)=\frac{32\pi K_{F}}{\left[(p-k)^{2}+m^{2}\right]^{2}},\label{eq: G_c}
\end{equation}

\begin{equation}
G_{g}(p-k)=\frac{3C_{2}\bar{g}^{2}\left(p-k\right)}{(p-k)^{2}+m_{g}^{2}\left(k^{2}\right)},\label{eq: G_g}
\end{equation}

where $K_{F}$ is the string tension, $m$ the mentioned entropic
parameter, $C_{2}$ the Cassimir eigenvalue, $m_{g}(k)$ the dynamical
gluon mass and $\bar{g}(k)$ the effective charge given by \cite{Cornwall1982}
\begin{equation}
\bar{g}^{2}\left(k^{2}\right)=\frac{1}{b\ln\left[\frac{k^{2}+4m_{g}^{2}\left(k^{2}\right)}{\Lambda_{QCD}^{2}}\right]},\label{eq:Effect Charge}
\end{equation}

where $b=(33-2n_{f})/48\pi^{2}$ is the one-loop coefficient in the
beta function with $n_{f}$ flavors.

Equation (\ref{eq:Comp GE}) can be simplified using the angular approximations
(discussed in \cite{Doff2011}, \cite{Cornwall(gMass)2008} and \cite{Cornwall2011}),
so that

\begin{equation}
f(x)=\int_{0}^{\infty}dy\left[F(x)\theta(x-y)+(x\leftrightarrow y)\right]\frac{yf(y)}{y+f^{2}(y)},\label{eq:Complete GE f(x)}
\end{equation}

with 
\begin{equation}
F(x;\rho,\gamma)=\left\{ \frac{ag(x)}{\left[x+\gamma\tilde{m}_{g}^{2}(x)\right]}+\frac{\rho}{\left(x+\rho/\rho_{c}\right)^{2}}\right\} ,\label{eq:F(x) Complete}
\end{equation}
 and 
\begin{equation}
g(x)=\ln^{-1}\beta\left[x+4\gamma\tilde{m}_{g}^{2}(x)\right].\label{eq:Effect Charge 2}
\end{equation}

Here we have used the definitions: $x=p^{2}/M^{2}$, $y=k^{2}/M^{2}$,
$f(x)=M(x)/M$, $g(x)=b\bar{g}^{2}(x)$, $a=3C_{2}/16\pi^{2}b$, $\beta=M^{2}/\Lambda_{QCD}^{2}$,
$\rho=2K_{F}/\pi M^{2}$, $\rho_{c}=2K_{F}/\pi m^{2}$, $m_{g}=m_{g}(0)$,
$\tilde{m}_{g}(x)=m_{g}(x)/m_{g}$ and $\gamma=m_{g}^{2}/M^{2}$.

Equation (\ref{eq:Complete GE f(x)}) can be transformed into the
boundary value problem 
\begin{equation}
\begin{array}{c}
F'(x)f''(x)-F''(x)f'(x)-\left[F'(x)\right]^{2}\frac{xf(x)}{x+f^{2}(x)}=0\\
\\
f(0)=1\qquad\mathrm{and}\qquad f'(0)=0,
\end{array}\label{eq:Complete GE BVP}
\end{equation}

with an extra IR condition (given by equation (\ref{eq:Complete GE f(x)}))%
\footnote{Here we are using $m_{g}=2\Lambda_{QCD}$ so that $\beta\gamma=4$.%
} 
\begin{equation}
1=\int_{0}^{\infty}dyF(y;\rho,\gamma)\frac{yf(y)}{y+f^{2}(y)}=I(\rho,\gamma).\label{eq:IR Cond Complete GE}
\end{equation}

The values of $(\rho,\gamma)$ which satisfy the previous condition,
correspond to the bifurcation points of the integral equation (\ref{eq:IR Cond Complete GE})
and with them we can find the dynamical quark mass $M$.

In the successive sections, we solve the boundary-value problem (\ref{eq:Complete GE BVP})
using the IR condition (\ref{eq:IR Cond Complete GE}) and also considering
the entropic condition $m\approx M$ in four cases: only with the
one-gluon sector; first with constant and then with running gluon
mass; only with the confining sector; and finally with both contributions.

\section{Fermionic gap equation with dynamical gluon mass\label{sec:Fermionic-gap-equation mg}}

With no confining contributions, the problem (\ref{eq:Complete GE BVP})
basically stays the same. The only difference is that now

\begin{equation}
F(x;\rho,\gamma)\equiv F_{g}(x;\gamma)=\frac{ag(x)}{\left[x+\gamma\tilde{m}_{g}^{2}(x)\right]},\label{eq:F(x) 1 gluon GE}
\end{equation}

and the IR condition reads %
\footnote{Where the subscripts $g$ are only to identify that the gap equation
has only the gluon contribution.%
}

\begin{equation}
1=\int_{0}^{\infty}dyF_{g}(y;\gamma)\frac{yf_{g}(y)}{y+f_{g}^{2}(y)}=I_{g}(\gamma).\label{eq:IR cond 1gluon GE}
\end{equation}

\subsection{Constant gluon mass}

\begin{figure}[h]
\includegraphics[scale=0.65]{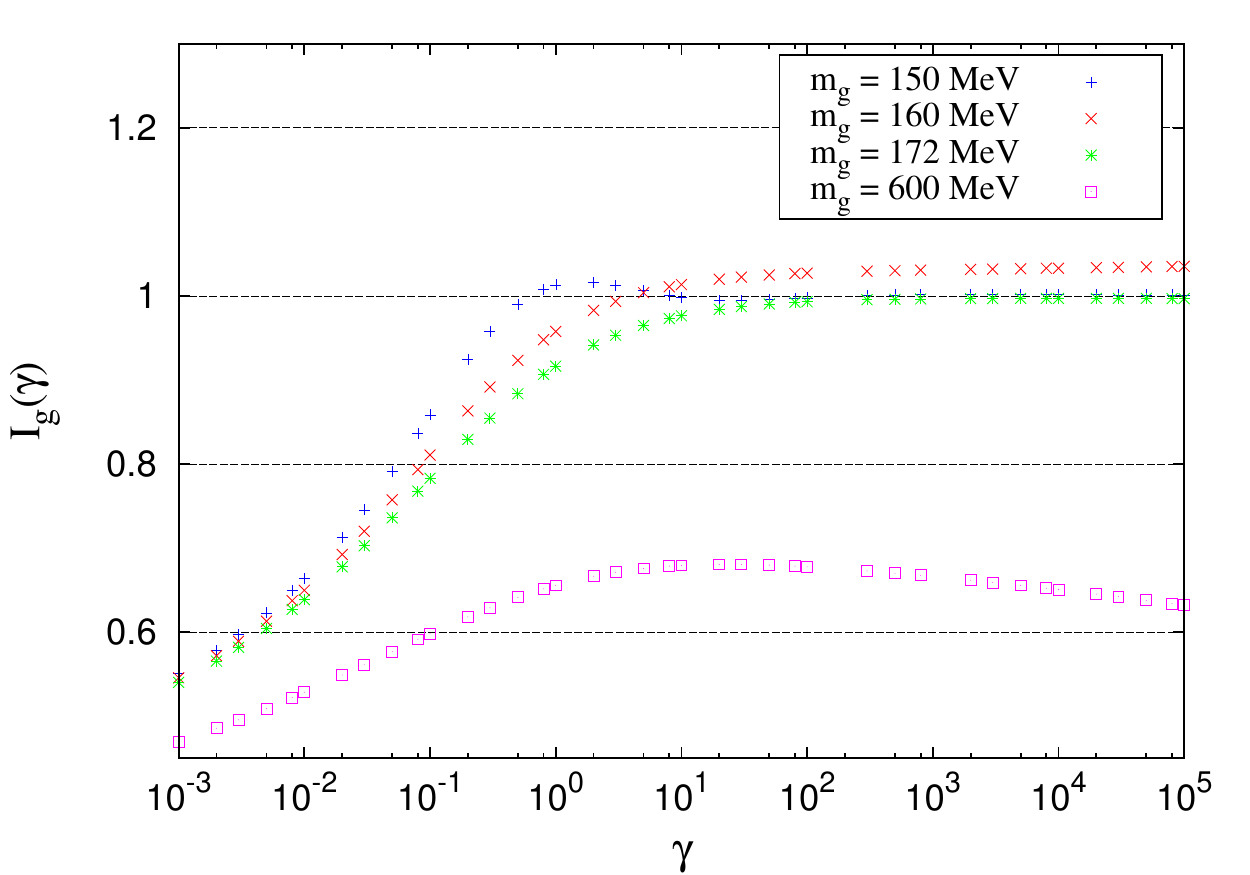} \caption{\label{fig: IR 1gluon}IR condition (\ref{eq:IR cond 1gluon GE})
for $n_{f}=2$, $\Lambda_{QCD}=300$MeV and different values of $m_{g}$.}
\end{figure}

In the case of a nonrunning dynamical gluon mass ($\tilde{m}_{g}(x)=1$),
the IR condition (\ref{eq:IR cond 1gluon GE}) is not satisfied for
the accepted phenomenological value $m_{g}\approx600\mathrm{MeV}$
(FIG. \ref{fig: IR 1gluon}). This situation is modified when considering
lower values for $m_{g}$. For example, we can see that for values
lower than $172$MeV, the IR condition starts to be satisfied, so
a condition for CSB to occur is $m_{g}\lesssim172$MeV. Another constraint
for the gluon mass is if $m_{g}<150$MeV, in which case the solution
of the problem (\ref{eq:Complete GE BVP}) starts to be unbounded
(because of the divergent values of the effective charge given at
low momenta). So, the dynamical chiral symmetry breaking for quarks
in the fundamental representation is constrained to the values for
the dynamical gluon mass (with the corresponding dynamical quark mass)
given by

\begin{equation}
\begin{array}{c}
150\lesssim m_{g}({\rm MeV})\lesssim172,\\
\\
190\gtrsim M({\rm MeV})\gtrsim0,
\end{array}\label{eq:Limits mg}
\end{equation}

\begin{figure}[h]
\includegraphics[scale=0.275]{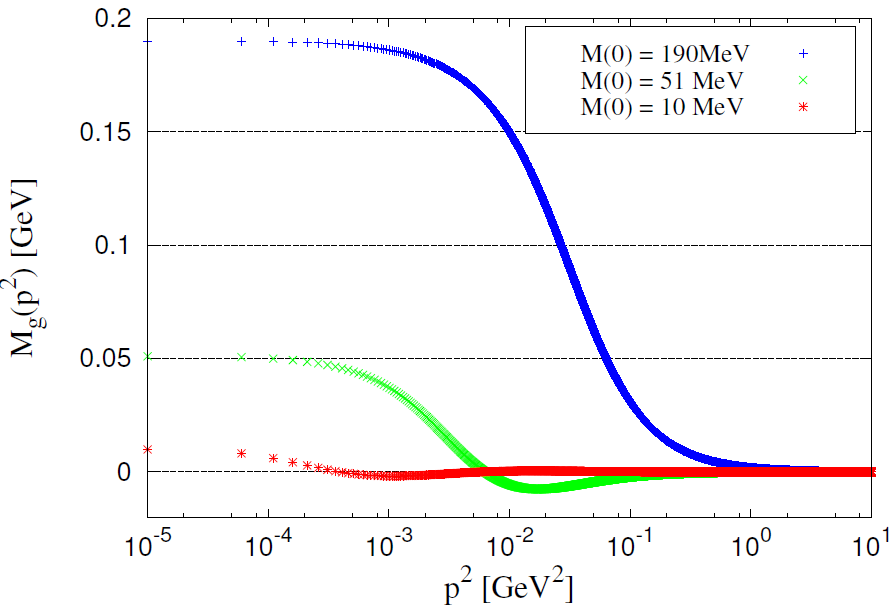} \caption{\label{fig: Mg(x)}Dynamical quark mass as a function of the momentum
for the fundamental vacuum and two replicas. This function is the
solution of the fermionic gap equation (\ref{eq:Comp GE}) with no
confining propagator ($G_{c}\rightarrow0$) and using $n_{f}=2$ and
$m_{g}=150$MeV.}
\end{figure}

For values within this condition, we can see how the multiple vacuum
states (the fundamental plus excitations) starts to appear. These
correspond to the points in which the condition $I_{g}(\gamma)=1$
is satisfied. For example, for $m_{g}=160\mathrm{MeV}$ we find only
one vacuum state, however for a value like 150MeV, there are three
which correspond to $\gamma=0.623,\,8.797,\,207.3$ and $M=190,\,51,\,10$MeV.
The solutions of the fermionic gap equation \ref{eq:Comp GE} (with
$G_{c}\rightarrow0$) which corresponds to the vacuum state and the
two replicas are shown in FIG. \ref{fig: Mg(x)} for $m_{g}=150\mathrm{MeV}$.
As we can see, the solutions can be classified by the number of zeros
as is done in \cite{Raya (Roberts)2013}.

\subsection{Running gluon mass}

The dynamical gluon mass term has the form \cite{Gonzalez2011}

\begin{equation}
\tilde{m}_{g}^{2}\left(k^{2}\right)=\left[\ln\left(\frac{k^{2}+\mu m_{g}^{2}}{\Lambda_{QCD}^{2}}\right)/\ln\left(\frac{\mu m_{g}^{2}}{\Lambda_{QCD}^{2}}\right)\right]^{-1-\delta},\label{eq: m_g(k^2) Cornwall 82}
\end{equation}

where $m_{g}^{2}$, $\mu$ and $\delta$ are parameters whose values
are chosen to fit the lattice data. We are going to use the phenomenological
values $m_{g}=2\Lambda_{QCD}$, $\mu=4$ and $\delta=1/11$.

This time, the condition for CSB is: 
\begin{equation}
\begin{array}{c}
177\lesssim m_{g}({\rm MeV})\lesssim204,\\
\\
310\gtrsim M({\rm MeV})\gtrsim0.
\end{array}\label{eq: Limits mg 2}
\end{equation}

\begin{figure}[h]
\includegraphics[scale=0.65]{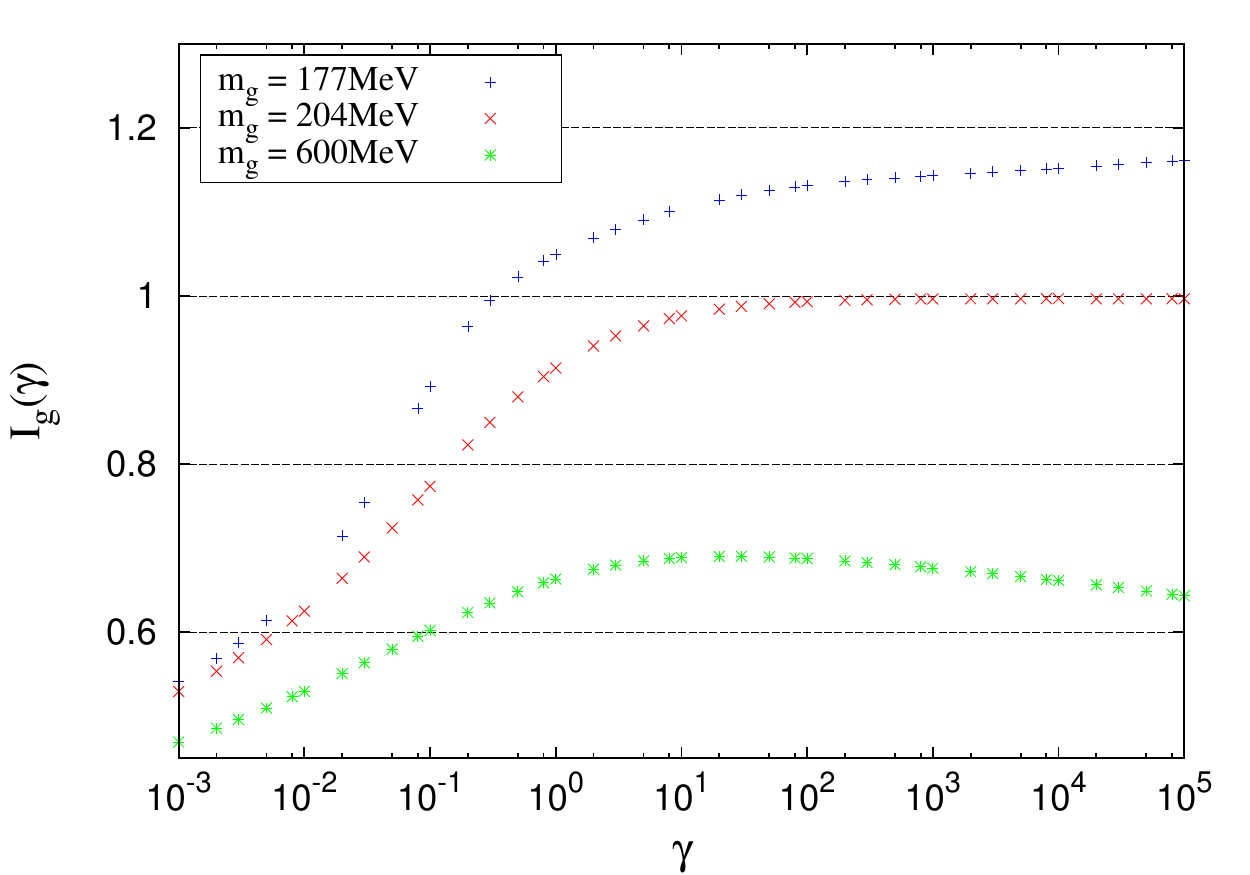} \caption{\label{fig: IR 1gluon run}IR condition with running gluon mass ($n_{f}=2$
and $\Lambda_{QCD}=300$MeV).}
\end{figure}

Within this interval, the gap equation has a single solution as we
can see in FIG. \ref{fig: IR 1gluon run} where the function $I_{g}(\gamma)$
intercepts the line $I_{g}=1$ (i.e. the IR condition (\ref{eq:IR cond 1gluon GE})
is satisfied) at exactly one point. This result is our basis for the
statement that when considering the running in the gluon mass the
replicas are absent in the limits of $m_{g}$ where the chiral symmetry
is broken. In this case, the solution of the gap equation has exactly
the same shape as in FIG. \ref{fig: Mg(x)} for $M(0)=190$MeV, but
now $M(0)=310$MeV.

\section{Fermionic gap equation with confining propagator\label{sec:Fermionic-gap-equation mc}}

In this case 
\begin{equation}
F(x;\rho,\gamma)\equiv F_{c}(x;\rho)=\frac{\rho}{\left(x+\rho/\rho_{c}\right)^{2}},\label{eq: F(x,rho) conf}
\end{equation}

so this time the free parameter in the IR condition is $\rho$ and
\begin{equation}
1=\int_{0}^{\infty}dyF_{c}(y;\rho)\frac{yf_{c}(y)}{y+f_{c}^{2}(y)}=I_{c}(\rho).\label{eq:IR conf}
\end{equation}

\begin{figure}[h]
\includegraphics[scale=0.65]{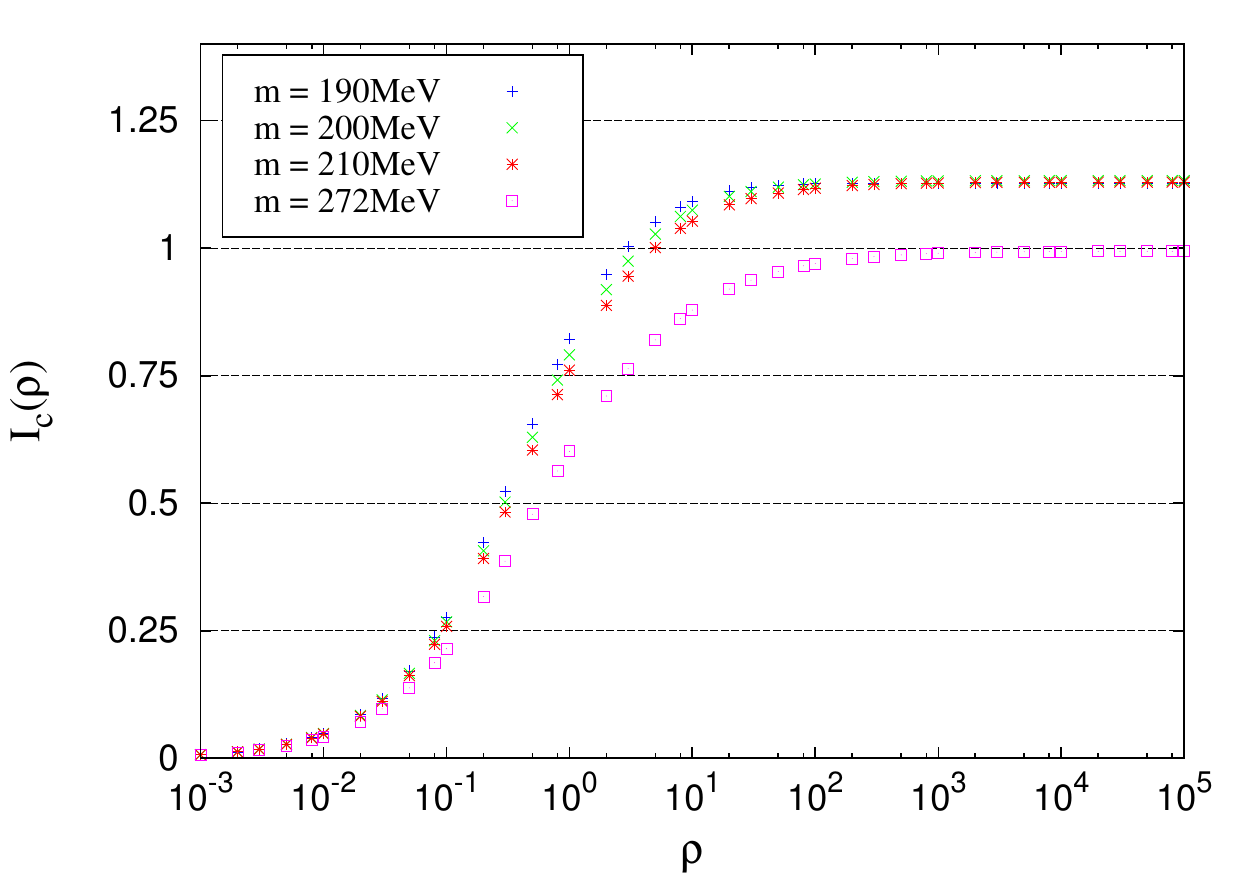} \caption{\label{fig: IR conf}IR condition for the confining equation with
$K_{F}=0.21$GeV$^{2}$ and different values of $m$.}
\end{figure}

For values of $m$ higher than $272$MeV the IR condition is not satisfied,
so that we can say that there is no CSB in that regime. For values
below this limit, we see (FIG. \ref{fig: IR conf}) how the curve
intercepts the line $I_{c}(\rho)=1$ at exactly one point, meaning
that there is only one vacuum state as it was called before.

Around the value $m=200$MeV, the resulting dynamical mass $M$ seems
to be close to the value of $m$ (TABLE \ref{tab: m vs M alpha}).
The value for which both the IR condition (\ref{eq:IR conf}) and
the entropic condition are (best) satisfied is $m=196$MeV for which
$M=199$MeV.

\begin{table}[h]
\begin{centering}
\begin{tabular}{|c|c|c|c|}
\hline 
$m$ (MeV)  & $\rho$  & $M$ (MeV)  & $\alpha=m/M$\tabularnewline
\hline 
\hline 
185  & 2.614  & 226  & 0.82\tabularnewline
\hline 
190  & 2.933  & 213  & 0.89\tabularnewline
\hline 
\textbf{196}  & \textbf{3.391}  & \textbf{199}  & \textbf{0.98}\tabularnewline
\hline 
200  & 3.754  & 189  & 1.06\tabularnewline
\hline 
205  & 4.290  & 178  & 1.15\tabularnewline
\hline 
\end{tabular}
\par\end{centering}

\caption{\label{tab: m vs M alpha}Relation between the parameter $m$, $M$,
$\rho$ and $\alpha$.}
\end{table}

\begin{figure}[t]
\includegraphics[scale=0.275]{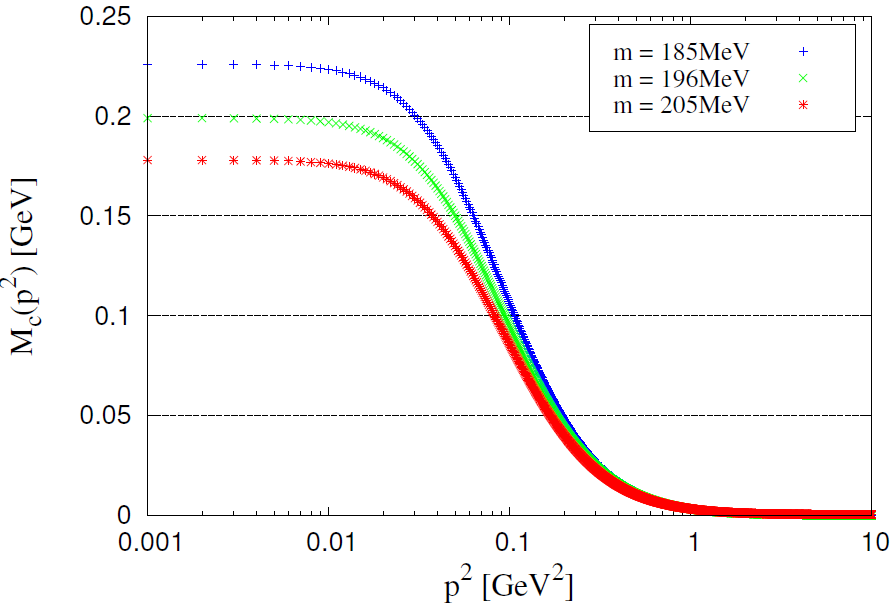} \caption{\label{fig: Mc(x)}Dynamical quark mass as a function of the momentum
for different values of the entropic parameter $m$. This function
is the solution of the fermionic gap equation (\ref{eq:Comp GE})
with no one-gluon-exchange propagator ($G_{g}\rightarrow0$) and using
$K_{F}=0.21$GeV.}
\end{figure}

To finish this section, we show the dynamical masses corresponding
to different values of $m=190,\,200\,{\rm and}\,210$MeV. We can see
(FIG. \ref{fig: Mc(x)}) how the solutions are bounded and well behaved
(non-oscillatory and non-negative in the domain).

\section{Fermionic gap equation - The complete case\label{sec:Fermionic-gap-equation comp}}

In our last case we are going to use the complete equation (\ref{eq:Complete GE f(x)}),
making use of (\ref{eq:F(x) Complete}), but with the relation $\gamma=\rho/\rho_{g}$
where $\rho_{g}=2K_{F}/\pi m_{g}^{2}$, so that $\rho$ is the only
free parameter in the IR condition. For the complete equation, the
situation turns out to be very similar to the one with the confining
equation. The values for the dynamical quark mass and the entropic
parameter are collected in TABLE \ref{tab: M complete ge}. The IR
condition is satisfied for only one value of the free parameter $\rho$,
so we can say that there are no vacuum replicas. The shape of the
function $I(\rho)$ is similar to the one shown in FIG. \ref{fig: IR 1gluon run}
(or \ref{fig: IR conf}). This time there are no constraints on the
dynamical gluon mass for CSB because the confining contribution is
driving most of the amount of the dynamical quark mass (as was noticed
in \cite{Cornwall2011}). The solutions for the gap equation $M(p^{2})$
are similar to those presented in FIG. \ref{fig: Mc(x)}, i.e. also
bounded and well behaved.

\begin{table}[h]
\begin{centering}
\begin{tabular}{|c|c|c|c|}
\hline 
$m$ (MeV)  & $\rho$  & $M$ (MeV)  & $\alpha=m/M$\tabularnewline
\hline 
\hline 
200  & 2.443  & 234  & 0.85\tabularnewline
\hline 
209  & 2.934  & 213  & 0.98\tabularnewline
\hline 
\textbf{210}  & \textbf{2.997}  & \textbf{211}  & \textbf{1.00}\tabularnewline
\hline 
211  & 3.062  & 209  & 1.01\tabularnewline
\hline 
220  & 3.743  & 189  & 1.16\tabularnewline
\hline 
\end{tabular}
\par\end{centering}

\caption{\label{tab: M complete ge}Dynamical mass for the complete gap equation.}
\end{table}

\section*{Summary and Conclusions}

We studied the fermionic gap equation with the inclusion of dynamical
massive gluons. We found that there is no CSB for the accepted value
of the dynamical gluon mass $m_{g}$. We also found limits on $m_{g}$
for dynamical chiral symmetry breaking in both cases; with constant
and running gluon mass. The former case shows the appearance of the
so called vacuum replicas, which correspond to higher order bifurcation
points in the gap equation. For the running case, the replicas are
absent in the limits where CSB is founded. We also studied the QCD-like
gap equation with the confining propagator proposed by Cornwall. In
this case we found CSB for values of the entropic parameter $m$ compatibles
with the entropic condition $m\approx M$. We finally studied the
complete gap equation which combines both contributions. In this last
case the replicas are also absent and CSB is present even for higher
values of the gluon mass compatibles with the phenomenology.

There are two possible scenarios for the relation between the chiral
symmetry restoration and the deconfinement phase: If confinement is
not necessary for chiral symmetry breaking, i.e. there is no need
for a confining propagator in the gap equation, the dynamical gluon
mass at zero momentum has to be constrained to the interval $177\lesssim m_{g}({\rm MeV})\lesssim204$.
Even more, we argue that it has to be closer to the lower limit to
obtain a dynamical quark mass according to phenomenology $m_{g}\approx180{\rm MeV}\rightarrow M\approx300{\rm MeV}$.
However, this result contradict the theoretical bound found in reference
\cite{Cornwall2009}, where the condition $m_{g}>1.2\Lambda_{QCD}$
is necessary to ensure the positivity of the imaginary part of the
gauge boson propagator. To clarify this scenario, simulations in the
lattice and accurate estimations of $m_{g}$ would be necessary. On
the other hand, if a confinement propagator is a necessary ingredient
into the gap equation, we can find CSB even for higher values of $m_{g}$
because confinement is driving most of the quark mass generation.
However, the confinement sector is not sufficient, because it is only
when we consider the complete gap equation that a good amount of CSB
is reached for the phenomenological value of the gluon mass ($m_{g}\approx600{\rm MeV}\rightarrow M=211{\rm MeV}$).

A final conclusion concerning the appearance of the multivacuum states
is that in any case, the non-perturbative effects (running gluon mass
or confinement) break the replicas and define a single vacuum, solution
of the fermionic gap equation.

\section*{Acknowledgments}

The author is thankful with D.McGee, J.Castro, A.A.Gomez and M.Varela
for their careful reading of this article. Further thanks to A.Martin
and A.Natale for discussions and advices about this manuscript, and
also to J.Cornwall for his reading and opinion of this document. This
research was partially supported by The University of Notre Dame,
by the Conselho Nacional de Desenvolvimento Cientifico e Tecnologico
(CNPq) and by the Coordenacao de Aperfeicoamento de Pessoal de Nivel
Superior (CAPES).

\end{document}